\documentclass[useAMS,usenatbib]{mn2e}
\usepackage{graphicx}
\title[The recognition of blazars and the blazar spectral sequence]
{The recognition of blazars and the blazar spectral sequence}
\author[S.~Ant\'on, IWA.~Browne]{S.~Ant\'on$^{1,2}$
\thanks{e-mail:Sonia.Anton@oal.ul.pt}, I.W.A. Browne$^1$\\
$^1$~Jodrell~Bank~Observatory, University of Manchester, Macclesfield, 
Cheshire, SK11 9DL, U.K.\\
$^2$~CAAUL, Observat\'orio Astron\'omico de Lisboa, Tapada de Ajuda, 1349-018, 
Lisboa, Portugal.}

\begin{document}

\date{}

\maketitle

\label{firstpage}

\begin{abstract}
We analyse a group of radio sources, a subset of the 200~mJy sample,
all of which have core-jet radio structures measured with VLBI and
have flat spectra stretching from the radio to the
millimetre/sub-millimetre band. Thus the objects have most
of the properties expected of blazars. However, they display varied
optical properties ranging from ``Seyfert-like'' objects, through BL
Lac objects, to ``normal'' elliptical galaxies. We investigate the
distribution of synchrotron peak frequencies in their Spectral Energy
Distributions (SEDs) and find a broad distribution between $10^{12}$
and $10^{16}$~Hz. Our conclusion is that we should consider virtually
all objects in the sample as blazars since much of the diversity in
their classification based on traditional optical criteria arises from
differences in the frequency at which the non-thermal emission begins
to decline.  Specifically, an object is only classified as BL Lac when
its peak frequency falls in the near IR/optical range. We
determine peak frequencies using the same method for objects from
other blazar samples. An important result is that our objects do not
follow the blazar spectral sequence proposed by Fossati et al. and
Ghisellini et al. in which, on average, peak frequencies increase as 
the radio luminosity decreases. Most of our low radio-luminosity sources 
have peaks in their SEDs at low frequencies, not at the expected high
frequencies. We suggest that at least part of the systematic trend
seen by Fossati et al. and Ghisellini et al. results from selection
effects.
\end{abstract}


\section{Introduction}
Blazars are radio sources that have prominent synchrotron cores.
Their broad-band spectra are smooth and flat from the radio up to the
infrared/optical bands (or beyond that), the emission is variable,
sometimes on time-scales as short as minutes to hours, and is often
polarised (see e.g. Urry \& Padovani (1995) and Kollgaard (1994) for
reviews). The blazar population comprises BL Lacs and Flat Spectrum
Radio Quasars, which are, respectively, featureless optical spectrum
objects and strong broad emission line objects. It is believed that
the continuum radiation of blazars is predominantly non-thermal
emission from a relativistic jet pointing to the observer at a small
angle to the line of sight, as proposed by Blandford \& Rees
(1978). Traditionally, apart from the radio-loudness, core-jet
morphology and flatness of the broad-band spectrum, an object is only
classified as a BL Lac if it has a strong non-thermal optical
component relative to the starlight and weak emission lines. The
emission lines are commonly quantified by the equivalent width (EW) of
the brightest line and the strength of the non-thermal optical
component is measured through the relative depression of the continuum
at 4000~\AA, the 4000~\AA\ break contrast, C. The usual limits adopted
are EW~$<$~5~\AA\ and C~$<$~0.25 (e.g. Stocke et al., 1991).  These
are empirical limits, based on the properties of a relatively small
number of objects. Results from newer samples have been showing that
the 4000~\AA\ break contrast distribution of BL Lacs and other
``non-BL Lac'' objects does not reveal a clear bi-modality
(e.g. March\~a et al., 1996; hereinafter M96, Laurent-Muehleisen et
al., 1998; Caccianiga et al., 1999; Rector et al., 2000).  As first
noted by M96 the traditional classification of a BL Lac is almost
certainly too restrictive, and an ``expanded'' definition that
includes objects with C $<$ 0.4 and objects with larger emission
line equivalent widths was proposed. More recently, Landt et al. (2004)
have suggested that there is a natural separation between weak- and
strong-lined sources in the [OIII] 5007~\AA/[OII] 3727~\AA\ plane which
clarifies blazar classification schemes.  Their weak-lined class
contains the majority of classical BL Lacs while the strong-lined
class has the majority of the broad-line blazars. Their approach is
interesting but might be complicated by the fact that they include a
wide range of luminosities in their sample. In this paper we look at
continuum spectral energy distributions (SED) of objects in a sample
with a restricted range of luminosities.\\

\noindent The SEDs  of blazars 
are distinctive by having a 2-hump shape, in
units of $\nu$ F$_\nu$ vs. $\nu$ (F$_\nu$ the flux density and $\nu$ the
frequency).  In general, the lower frequency hump is attributed to
synchrotron radiation and the second hump to inverse Compton
scattering by the synchrotron electrons. The first hump is
characterised by the frequency $\nu_{peak}$ at which the synchrotron
component begins to decline. This frequency is directly related to a
break in the synchrotron electron energy spectrum in the jet. Some BL
Lacs (and Flat Spectrum Radio Quasars) show the first energy peak in
the infrared/optical wavelength range and these are often labelled Low
frequency peaked BL Lacs (LBLs). On the other hand
there are those that show the first energy peak in the EUV region, the
high frequency peaked BL Lacs (HBLs). LBLs are mainly
found in the radio-selected samples, whereas the HBLs are mainly found
in the X-ray selected samples. Although some authors disagreed
(e.g. Polatidis, 1989), for some time it was thought that there
was a bi-modal LBL-HBL population.  However, the discovery of
more BL Lacs, mainly in the X-ray/radio surveys, has been showing that
there is a continuous distribution of the peak frequencies
(e.g.  Perlman et al. 1998, 2001) and that the
apparent dichotomy was due to selection effects (e.g. Padovani et al.
(2003) and references therein). It is now widely accepted that BL Lacs
have a broad $\nu_{peak}$ distribution, spanning $\sim$ 5 orders of
magnitude in frequency (10$^{13}$~$<\nu_{peak}$~$<$~10$^{18}$~Hz), the
extremes being populated by LBLs and HBLs.  \\

\noindent A potentially very important result has been pointed out by Fossati 
et al.  (1998) and Ghisellini et al. (1998, 2002). They found a sequence
involving the frequency at which the peak of the non-thermal emission
occurs and the radio luminosity; the synchrotron peak position and the
radio luminosity appear anti-correlated in the sense that lower peak
positions are associated with the more luminous radio emitters. Donato
et al. (2001), based on X-ray spectra of blazars, also found the same
anti-correlation. This recognition of a blazar spectral sequence
appears to be an important breakthrough in our understanding and it is
therefore vital to check if the result holds when a wider range of
parameter space is explored.  \\

\noindent Much of the research on blazars has been
based on very bright and relatively small samples, either selected on
the radio band, e.g.  ``1 Jy sample'' (Stickel et al., 1991) or in the
X-ray band, e.g.  EMSS\footnotemark\footnotetext{Einstein Extended
Medium Sensitivity Survey} (Gioia et al., 1990; Stocke et al., 1991;
Maccacaro et al., 1994). As a consequence, the knowledge on these
objects has been built on the properties of the most powerful ones,
which may not be representative of the population as a whole. It would
be, therefore, very useful to have a new radio-selected sample of
relatively low radio-luminosity objects in order to investigate
further the luminosity-peak frequency sequence. In particular, if the
trend in the data is telling us about the physical properties of the
blazars, and not about selection effects, the low-luminosity objects
should be found to have relatively high synchrotron peak
frequencies. Such an investigation is particularly timely as
recent results point to a more complicated scenario, in which objects
from deep surveys do not follow the anti-correlation trend (Padovani
et al., 2003).\\
 In order to investigate this issue and others we have
been analysing a low-luminosity radio-selected sample, the 200~mJy
sample (Ant\'on et al, 2004, hereinafter A04; M96).  
Producing statistically well-defined samples of blazars
is difficult since traditionally it has relied on the detection of an
optical non-thermal component in the SED. This introduces strong
luminosity-dependent selection effects (Browne \& March\~a, 1993;
March\~a \& Browne, 1995, Landt et al. 2002, 2004; Rector \& Stocke, 2001) 
and many analyses of blazar population
statistics can be suspect for this reason. Therefore we have been
exploring an approach to sample selection focusing on the low
frequency (radio, mm and sub-mm) properties, particularly the
possession of milliarcsecond VLBI core-jet structure of the type
commonly found in superluminal radio sources.  With our 200 mJy 
core-jet sample we have a set of 
objects well suited to extend our knowledge of the SEDs of blazar-like 
objects. Here we address the following questions:
\begin{itemize}
\item What is the range of synchrotron peak frequencies amongst the
 low-luminosity radio selected objects? Do these objects follow the 
 luminosity peak frequency sequence? 
\item How does the peak frequency influence the classification of the objects?
\item Are there any selection effects influencing the luminosity peak 
frequency  sequence?
\end{itemize}
The paper is organised as follows. In section~\ref{sec-seds} we present a
summary of the properties of the sources discussed here, as well as
the analysis of their SEDs.  In section~\ref{sec-peak} we discuss the
relation between the SED type and the optical classification of the
objects. The properties of the 200 mJy objects in the framework of the
spectral sequence found by Fossati et al. (1998) and Ghisellini 
et al. (1998, 2002) are analysed in
section~\ref{sec-specseq}. Specifically, we fit a model consisting
of a low frequency power law plus a higher frequency parabola to both
the 200~mJy SEDs and those from other samples in order to look for
systematic trends. The main points of this work are summarised in
section~\ref{sec-conclusion}.  Throughout the paper we assume
H$_o$ = 65 kms$^{-1}$ Mpc$^{-1}$ and $q_o$=0.

\section{The SEDs}
\label{sec-seds}

\begin{table}
\begin{tabular}[t]{lccccc}
\hline \hline
\multicolumn{1}{c}{Object} & \multicolumn{1}{c}{Radio} &
\multicolumn{1}{c}{z} & \multicolumn{1}{c}{Optical}  & 
\multicolumn{1}{c}{SED} & \multicolumn{1}{c}{T$_{\mbox {\tiny B}}$} \\
\multicolumn{1}{c}{B1950} & \multicolumn{1}{c}{morph.} & 
\multicolumn{1}{c}{} & \multicolumn{1}{c}{class} & 
\multicolumn{1}{c}{type} &  \multicolumn{1}{c}{$\times 10^{9}$ K} \\
\hline
0055+300 & c+j$^{14}$  & 0.015 & PEG & BPL+IB  &  6.2 $^{1,a}$\\
0109+224 & c+j$^{13}$  &       & BLL & BPL     & 12.2 $^{2,b}$ \\ 
0125+487 & c+j$^{1}$   & 0.067 & Sy1 & BPL     &  1.5 $^{2,b}$ \\
0149+710 & c+j$^{1}$   & 0.022 & PEG & BPL+IB  &  2.4 $^{2,b}$ \\
0210+515 & c+j$^{1}$   & 0.049 & BLL & BPL     &  0.8 $^{2,b}$ \\
0251+393 & c+j$^{7}$   & 0.289 & Sy1 & BPL     & 15.0 $^{3,b}$ \\
0309+411 & c+j$^{7}$   & 0.134 & Sy1 & BPL     & \\
0321+340 & c+j$^{9}$   & 0.061 & Sy1 & BPL     & \\
0651+428 & c+j$^{1}$   & 0.129 & BLC & BPL     &  1.0 $^{2,b}$ \\
0716+714 & c+j$^{3}$   &       & BLL & BPL     &  4.6 $^{1,a}$ \\
0806+350 & c+j$^{2}$   & 0.082 & BLL & BPL     &  0.8 $^{2,b}$ \\
0912+297 & c+j$^{13}$  &       & BLL & BPL     & \\
1055+567 & c+j$^{1}$   & 0.144 & BLL & BPL     & 2.4 $^{2,b}$ \\
1101+384 & c+j$^{11}$  & 0.031 & BLL & BPL     & 7.8 $^{1,a}$ \\
1123+203 & c+j$^{9}$   & 0.133 & BLL & BPL     & \\
1133+704 & c+j$^{8}$   & 0.046 & BLL & BPL     &\\
1144+352 & c+j$^{7}$   & 0.063 & PEG & BPL+IB  &\\
1147+245 & c+j$^{4}$   &       & BLL & BPL     &\\
1215+303 & c+j$^{1}$   & 0.130 & BLL & BPL     & 4.2 $^{2,b}$ \\
1219+285 & c+j$^{5}$   & 0.102 & BLL & BPL     & 3.2 $^{1,a}$ \\
1241+735 & c+j$^{1}$   & 0.075 & PEG & BPL     & 1.8 $^{2,b}$ \\
1246+586 & c+j$^{12}$  &       & BLL & BPL     & 8.0 $^{4,b}$ \\
1418+546 & c+j$^{4}$   & 0.151 & BLL & BPL     & \\
1421+511 & c+j$^{9}$   & 0.274 & Sy1 & BPL     & \\
1424+240 & c+j$^{9}$   &       & BLL & BPL     & \\
1551+239 &             & 0.117 & PEG & BPL+IB  & \\
1646+499 & c+j$^{1}$   & 0.049 & Sy1 & BPL     & 1.4 $^{2,b}$ \\
1652+398 & c+j$^{10}$  & 0.030 & BLL & BPL     & 6.7 $^{1,a}$ \\
1658+302 &             & 0.036 & PEG & BPL+IB  &\\
1744+260 & c+j$^{15}$  & 0.147 & Sy2 & BPL     & \\
1807+698 & c+j$^{6}$   & 0.051 & BLL & BPL     & 13.2 $^{1,a}$ \\
1959+650 & c+j$^{1}$   & 0.047 & BLL & BPL     & 2.4 $^{2,b}$ \\
2116+81  & c+j$^{16}$  & 0.084 & Sy1 & BPL     & \\
2320+203 & c+j$^{2}$   & 0.038 & PEG & BPL     & 0.8 $^{2,b}$ \\
\hline \hline
\end{tabular}
\caption{
1st column 1950.0 IAU name of the object, 2nd column is the radio 
morphology, 3rd column is the optical classification,  4th column is the SED 
type, 5th column is the brightness temperature.
 {\it Radio morphology classification key}: ({\bf c+j}) core+jet VLBA structure
according to: 
 1. Bondi et al. (2001)
 2. Bondi et al. (2004)
 3. Gabuzda et al. (1998)
 4. Gabuzda et al. (1996)
 5. Gabuzda \& Cawthorne (1996)
 6. Gabuzda et al. (1989)
 7. Henstock et al. (1995)
 8. Kollgaard et al. (1996)
 9. NRAO VLBA calibrators list
10. Pearson et al. (1993)
11. Piner et al. (1999)
12. Taylor et al. (1994)
13. USNO Astrometric VLBI database
14. Venturi et al.(1993)
15. The object 1744+260 shows core-jet morphology at MERLIN resolution 
     (Augusto et al., 1998). 
16. Taylor et al. (1996).
{\it Optical classification keys}: 
{\bf BLL}=BL Lac, {\bf BLC}=BL Lac Candidate, {\bf Syf1}=Seyfert-1 type, 
{\bf Syf2}=Seyfert-2 type, {\bf PEG}=Passive Elliptical 
Galaxies. {\it SED classification keys}: {\bf BPL+IB}  broadband flat spectra 
with one or more bumps, {\bf BPL} broken power-law broadband spectra.
The brightness temperatures where computed from VLBI maps,
at $^{a}$  15 GHz or $^{b}$  5 GHz presented in:
1. Kellermann et al. (1998) 
2. Bondi et al. (2004) 
3. Henstock et al. (1995) 
4. Taylor et al. (1994)
}

\label{tabela}
\end{table}

\subsection{The sample of objects}
\label{sec-objects}
The objects analysed in this paper are a subsample of the 200~mJy
sample (A04, M96). This is a nearby radio-selected sample (most of the
objects have z$<$ 0.2), that comprises objects with core-dominated
morphology at 8.4 GHz, with S$_{\mbox{\tiny 5GHz}}$ $\geq$ 200 mJy,
flat radio spectra between 1.4 and 5 GHz 
($\alpha_{1.4}^{5.0}$ $\geq-$ 0.5) and R $\leq$ 17 mag.\\

\noindent The objects under consideration in the present work are presented in 
Table~\ref{tabela}. They were selected on the basis of the results from 
the multiwavelength study of the 200 mJy sample, presented in A04. That study
has revealed a subset of objects with uniform radio/mm properties, flat
broadband spectra, some being classified as BL Lac objects but some having 
different optical classifications, as follows:
\begin{enumerate}

\item From the analysis of the SEDs of the 200 mJy sample as a whole
(A04), it became clear that there was a subset of objects which had,
at least at wavelengths down to the sub-mm, flat broadband spectra (in
units of flux density vs frequency) before steepening, and they have
been classified as broken-power-law type objects ({\bf BPL}) in Table
1. There are a few objects showing an infrared bump that we interpreted as
dust emission, and they are listed in Table 1 as {\bf BPL+IB}
(broken-power-law + infrared bump)\footnotemark\footnotetext{see A04
for details; there the BPL+IB and the BPL are called type II and III
respectively.}.

\item All the objects have a {\bf core-jet} radio structure when
observed with VLBI techniques (Bondi et al 2004, 2001 and references
in Table 1).  This ensures that we are dealing with a set of objects
in which the radio emission is mainly from the active nucleus and not
from compact lobes, as would be the case with sources with CSO-like
radio structures. In addition we note that the possession of core-jet
VLBI radio structures is in general a good circumstantial evidence
that we are dealing with emission from relativistic jets. For example,
Britzen et al. (Britzen, personal communication) have made extensive
VLBI observations of a complete sample of 297 flat spectrum radio
sources (CJF: Taylor et al. 1996) and they find that $\sim$90\% of
those CJF sources with core jet structures show superluminal motion.\\
For most of the objects the brightness temperatures could be
estimated, and they are presented in Table 1. The temperatures were
computed according to: \\ \centerline{ T$_{\mbox B}$=
1.8$\times$10$^{9}$ $\frac{{\mbox S}_\nu {\mbox {\small [mJy]}}}
{\nu^2 \ {\mbox {\small [GHz]}} \ \ \Theta^2}$}
\noindent  with $\Theta$ being the source size
in milliarcsec (Caccianiga et al., 2001). Note that T$_{\mbox B}$ is a lower 
limit, as the size of 
the emitting region is most probably smaller than the beamsize. 
All the objects have brightness temperatures larger than $10^{8}$ K, showing
that the emission is certainly non-thermal in origin. More interesting is 
the fact that the T$_{\mbox B}$ of BL Lacs and of the other objects are very 
similar.

\item  Based on optical spectral properties of the objects, in particular
the  4000~\AA\ break contrast and
emission line strength, they are classified as (see
A04 and M96):\\ 
\noindent $\bullet$ BL Lac objects ({\bf BLL}), objects that obey the
``classical'' BL Lac definition: C~$\leq$~0.25 and EW~$\leq$~5~\AA.\\
$\bullet$ BL Lac Candidates ({\bf BLC}), objects with relatively small break
contrast and EW: 0.25~$<$~C~$<$~0.4 and EW~$<$~40~\AA. \\
$\bullet$ Seyfert-like objects, objects with relatively strong broad ({\bf
Sy1}) or narrow ({\bf Sy2}) emission lines: EW~$>$~40~\AA.\\ 
$\bullet$ Passive Elliptical Galaxies ({\bf PEGs}), objects with high contrast
 and weak emission lines: C~$\geq$~0.40 and EW~$<$~40~\AA.

\end{enumerate}

\subsection{Fitting the SEDs}
\label{sec-sedfitting}

\begin{figure*}
\centerline{
\includegraphics[width=7cm]{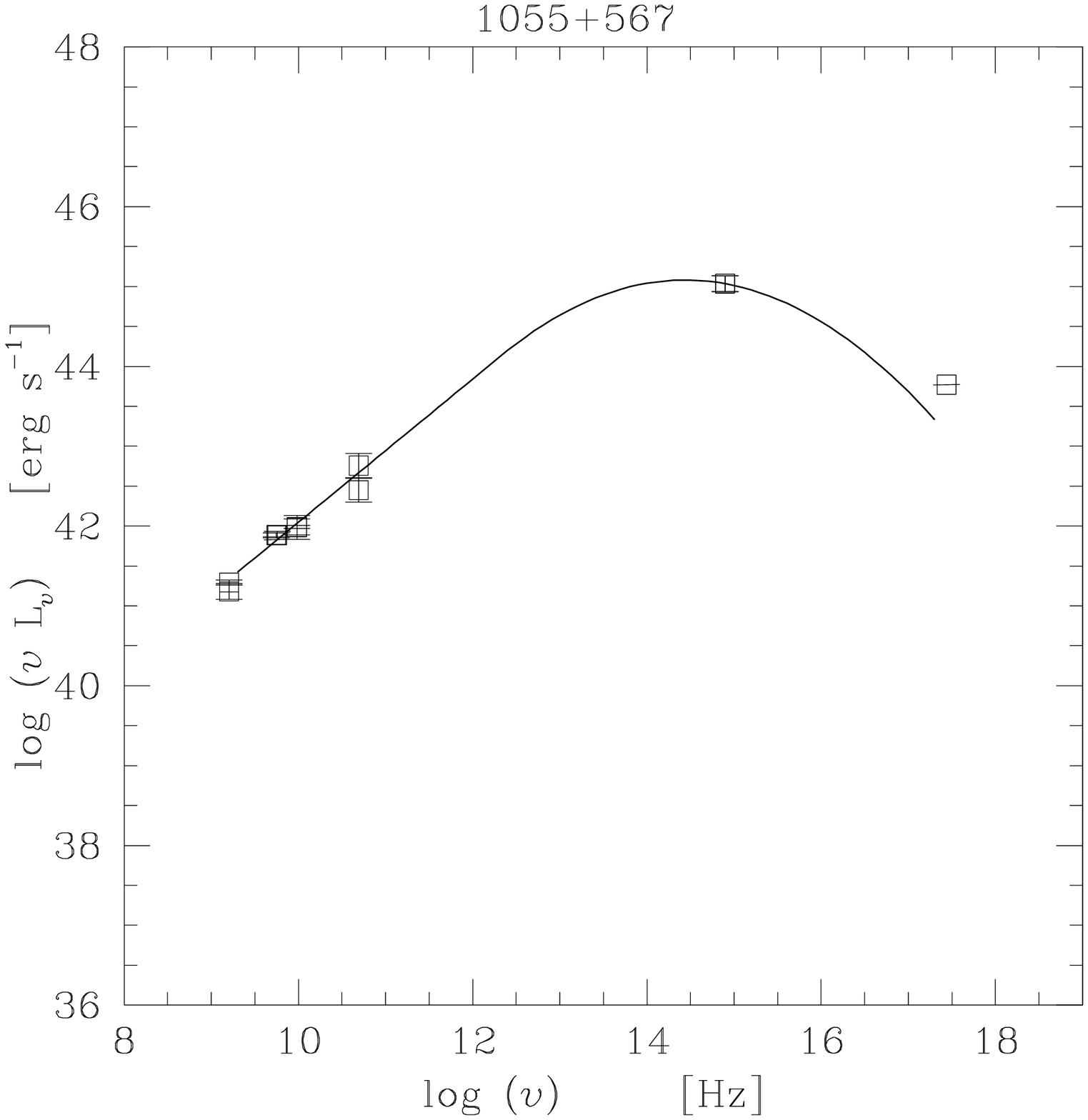}
\hspace{1.5cm}
\includegraphics[width=7cm]{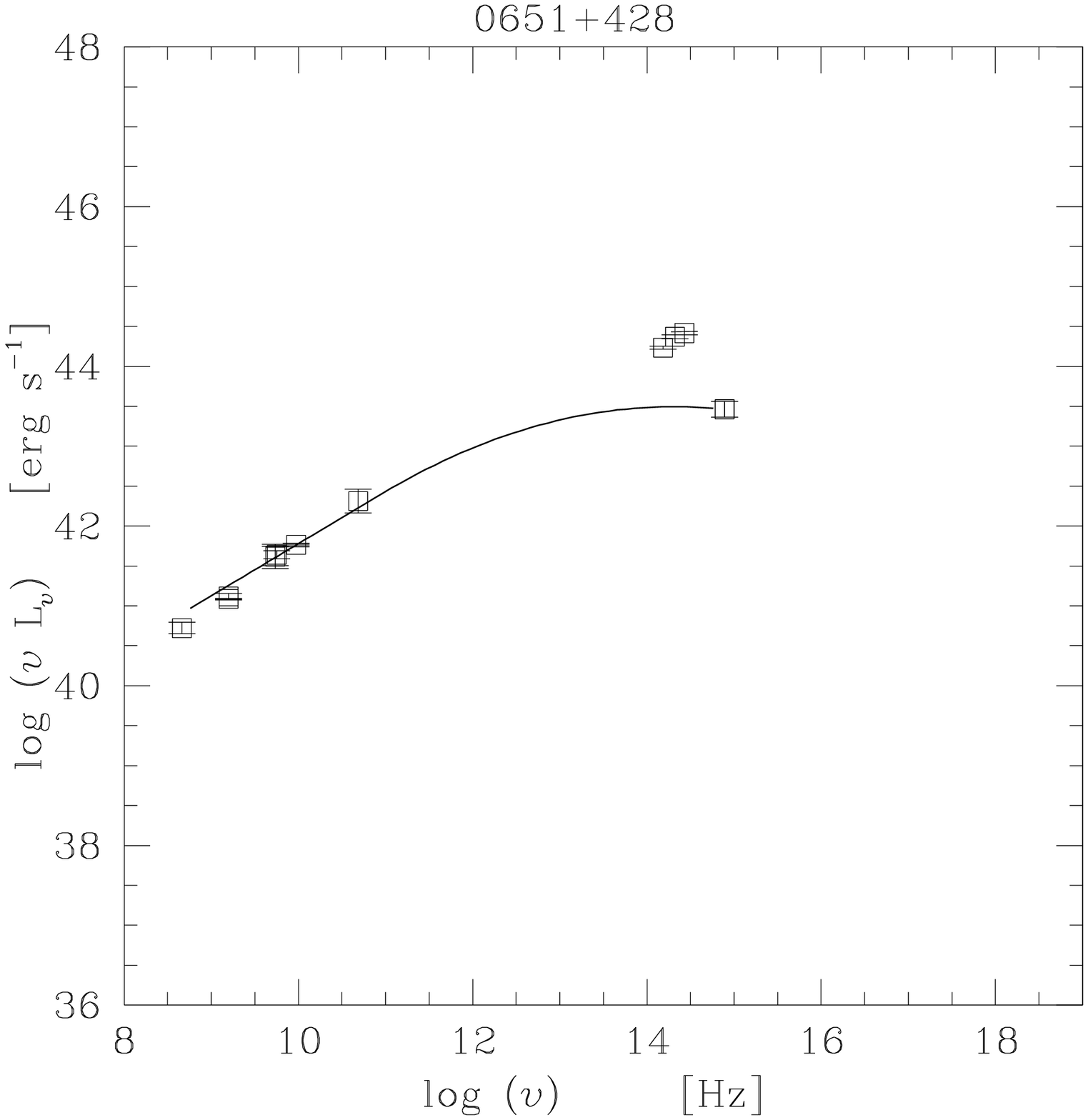}}
\centerline{
\includegraphics[width=7cm]{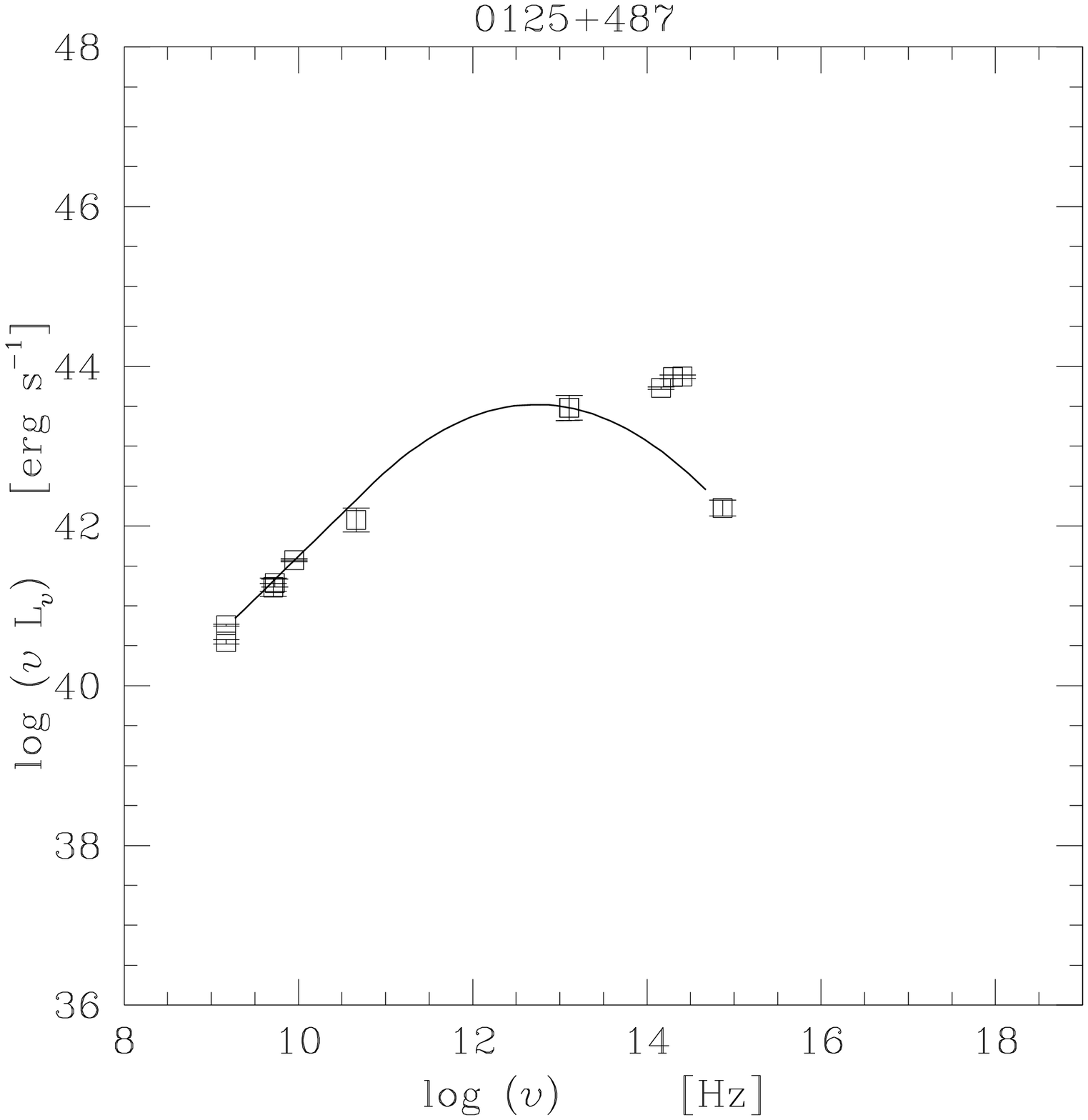}
\hspace{1.5cm}
\includegraphics[width=7cm]{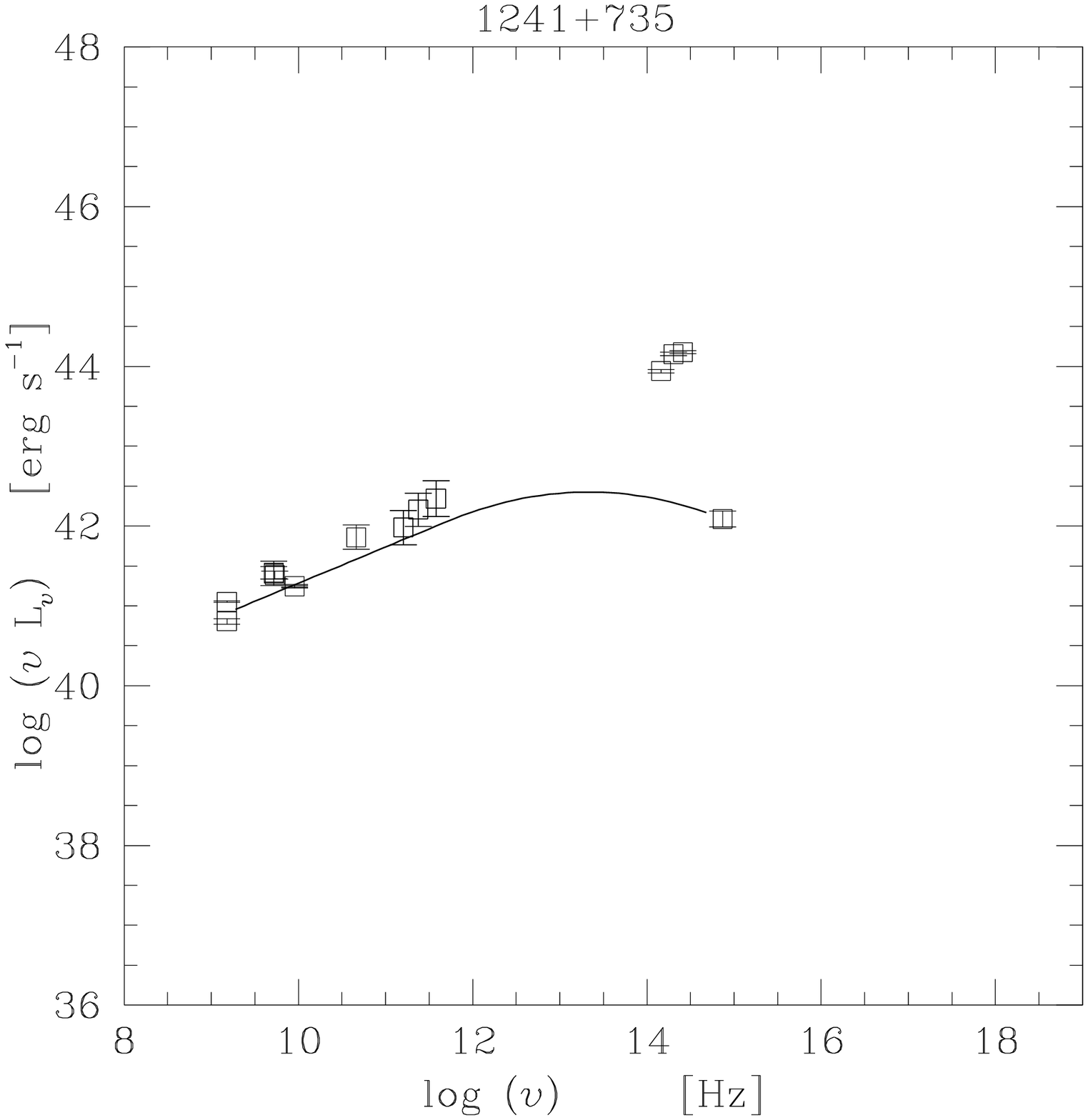}}
\caption{Spectral energy distributions, in units of log ($\nu$ L$_\nu$)
$vs$ log ($\nu$), of objects with different optical classification: BL
Lac object 1055+567, BL Lac Candidate object 0651+428, Seyfert-1 like
object 0125+487 and the PEG 1241+735.  The data are from A04. The
superimposed line is the fit of a power-law together with a parabolic
function (see the text).}
\label{figuras}
\end{figure*}
The multiwavelength study of the 200 mJy sample has revealed an
homogeneous subset of objects which have many of the properties
expected of BL Lacs but not all of which met the traditional BL Lac
selection criteria. As it can be inferred from Table~\ref{tabela} all
the objects that are optically classified as BL Lacs have SEDs which
 fall into the category BPL. More interesting is the fact that
objects other than BL Lacs look similar to BL Lacs up to relatively
high frequencies. To investigate this further, the SEDs of the objects
in Table~\ref{tabela} were model fitted to estimate the peak
frequency, and look for any statistical difference between BL Lacs and
non-BL Lac objects.  We assumed the Fossati et al. (1997) model, that
consists of a power-law, to represent the low frequencies, plus a
parabola for the higher frequencies. The model differentiates two
frequencies: up to a certain frequency the radiation is taken to be
optically thick and it is represented by the power law, then the
emission becomes optically thin and it is represented by the parabolic
function with the maximum occurring at a frequency $\nu_{peak}$.  More
complex and physically based models have been considered by others
(e.g. Padovani et al 2003). But, as in our case the aim is to compare
the peak frequencies amongst objects, it is sufficient to use a
convenient analytical function for the modelling.  The same protocol
was used for all the objects. The flux densities were k-corrected,
assuming the spectral indices from Sambruna et al. (1996) and in the
computation of the luminosities a redshift of 0.3 was
assumed\footnotemark\footnotetext{We assume 0.3 because, for lower
redshifts, it is often possible to resolve the host galaxy emission
from the nuclear emission and measure the redshift of the former.}  if
there was no redshift available for the object. When it was clear that
the near infrared/optical emission was predominantly starlight (not
AGN emission), the data points were ignored during the fitting
process. We note that the optical data point is our estimate of the
non-thermal emission, based on the 4000~\AA\ break contrast as
described in A04. Figure~\ref{figuras} shows the SEDs of four objects,
each representing a different optical class: 1055+567 is a BL Lac,
0651+428 is a BL Lac Candidate, 0125+487 is a Seyfert-1 like object
with blazar characteristics and 1241+735 is a PEG.

\section{Peak frequency \& Optical classification}
\label{sec-peak}

\begin{figure}
\includegraphics[width=8cm]{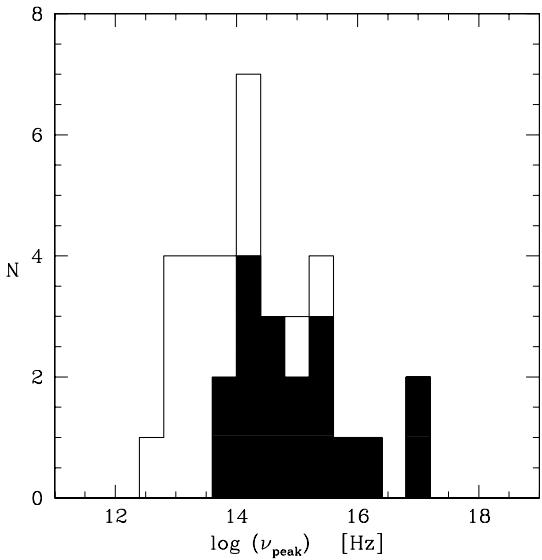}
\caption{Peak frequency distribution of the objects in Table~\ref{tabela}.
Black area represent the $\nu_{peak}$ distribution of known BL Lac objects}
\label{peaks}
\end{figure}

\begin{figure}
\includegraphics[width=8.5cm]{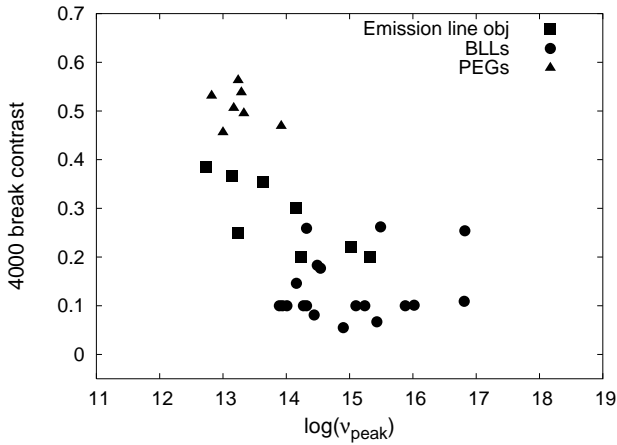}
\caption{4000~\AA\ break contrast against peak frequency. The PEGS are 
represented by square 
symbols, Sy-like objects are represented by stars, BL Lac Candidate objects 
by triangles and 
BL Lac objects are represented by circles.}
\label{cpeak}
\end{figure}

The SEDs were analysed in the manner described in the previous section
and the distribution of the synchrotron peak frequencies
($\nu_{peak}$) is shown in Figure~\ref{peaks}. We draw attention to
the fact that the $\nu_{peak}$ are widely and fairly {\it
continuously} distributed from $\sim$ 10$^{12}$ to 10$^{16}$ Hz with
the well-known BL Lacs (C~$<$~0.25 \& EW~$<$~5~\AA) and other objects
(C~$>$~0.25 \& EW~$>$~5~\AA) occupying  overlapping
regions. The $\nu_{peak}$ distribution strongly points to the
conclusion that we are dealing with a single group of objects drawn
from the same population (otherwise one would not expect a continuous
distribution), containing a wide spread of non-thermal cut-offs,
ranging over 4 orders of magnitude in frequency. \\
In the lower
frequency tail of the distribution, we suggest that we are seeing the
red end of the blazar sequence and, if one wished to maintain the
tradition of labelling objects by where their spectral peaks lie, one
might call the objects with $\nu_{peak}$ lying at frequencies smaller
than the near-infrared as Very Low-frequency-peaked BL Lacs, or
VLBLs. We do not, however, believe it helpful to make this
distinction and wish, rather, to emphasise the similarities. These
objects:\\
\noindent $\bullet$  all have flat broad-band spectra
 up to frequencies of $\sim$100~GHz or higher.\\
\noindent $\bullet$ all have radio structures that consist of
compact VLBI cores plus jets.\\ How do the values of $\nu_{peak}$
relate to the traditional optical classifications of BL Lacs,
Seyfert-like objects and PEGs? It is of course of interest that not
all the objects listed in Table~\ref{tabela}, and that we would like
to label as blazars, meet the usual ``BL Lac'' criteria.  In
Figure~\ref{cpeak} the values of $\nu_{peak}$ are plotted against the
4000~\AA\ break contrasts. As expected, the $\nu_{peak}$ distributions
of BL Lacs are different from those of Seyfert-like, BL Lac Candidates
and PEG objects; BL Lacs have $\nu_{peak}$ in the
near-infrared/optical/X-ray bands and the rest of objects have their
$\nu_{peak}$ located mainly between the long and near infrared bands.
Of course it should be no surprise that the strength of the
synchrotron component in the optical band (which is the basis of normal BL Lac 
classification) is dependent on the
frequency at which the synchrotron emission begins to decline. Note
that the importance of this effect on traditional classifications
decreases with increasing non-thermal luminosity. In the case of
powerful objects, their non-thermal optical component is strong enough
to compete with the starlight (from the host galaxy) and to fill the
4000~\AA\ break, even in the case of a relatively low
$\nu_{peak}$. That is, when there is a wide range of peak frequencies,
as we believe is the case, it can lead to a selection-induced
anti-correlation between $\nu_{peak}$ and luminosity (see below). This
implies that a very low luminosity BL Lac is only recognised as such
if it has a very high frequency peak, whereas for a high luminosity
object recognition is possible for a wide range of peak frequencies.

\section{The trend of peak frequency with radio luminosity}
\label{sec-specseq}

\begin{figure}
\includegraphics[width=8.5cm]{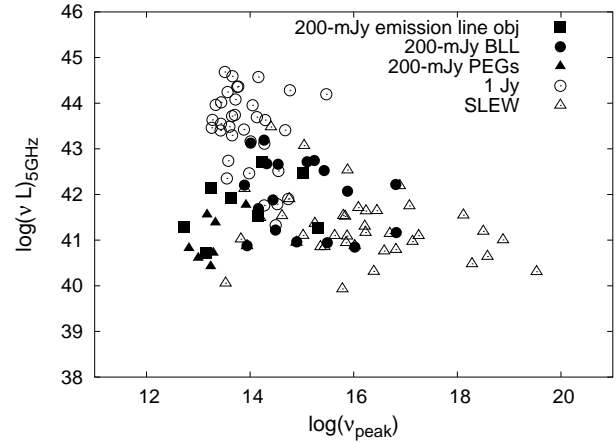}
\caption{5 GHz radio luminosity vs peak frequency for 1 Jy, Slew and 200 mJy 
objects. The latter are plotted in different symbols depending on whether
 they are known BL Lacs (BLLs), have emission lines or weak emission lines 
(PEGs)}
\label{lum5nupeak}
\end{figure}

A crucial question is whether or not the systematic shift of the
average synchrotron peak frequency with luminosity, pointed out by
Fossati et al. (1998) and Donato et al. (2001) is confirmed for our
200~mJy sample of low-luminosity radio-selected objects. This spectral
sequence is important because, if true, it represents a great
empirical simplification which is backed by an elegant physical
explanation in terms of different inverse Compton losses occurring
in the jet due to ambient photon densities which vary with object
luminosity (Ghisellini et al. 1998; Georganopoulos et al.,
2001). \\

\noindent Unfortunately there are also selection effects which might give
rise to a spurious correlation of the form that is observed 
(see also Perlman et al. 2001). This is
because the result is based on all available blazar SEDs taken from a whole
range of samples with different selection criteria. One of the
possible selection effects -- i.e. that recognition of BL Lacs with
low cut of frequencies is more difficult at low luminosities than high
luminosities -- was already pointed out in Section~\ref{sec-peak}.
Another possible selection effect is that in the samples analysed by
Fossati et al., the lower luminosity objects are mainly represented by
the X-ray selected objects (e.g. the Slew sample) and therefore likely to
be mainly HBL-type. On the other hand, the higher luminosity objects
represented by the 1 Jy and 2 Jy samples, are mainly radio-selected
objects, and therefore the majority are of LBL-type. Thus a
correlation of the form observed could arise from the type of objects
available for analysis. \\

\begin{figure*}
\includegraphics[width=18cm]{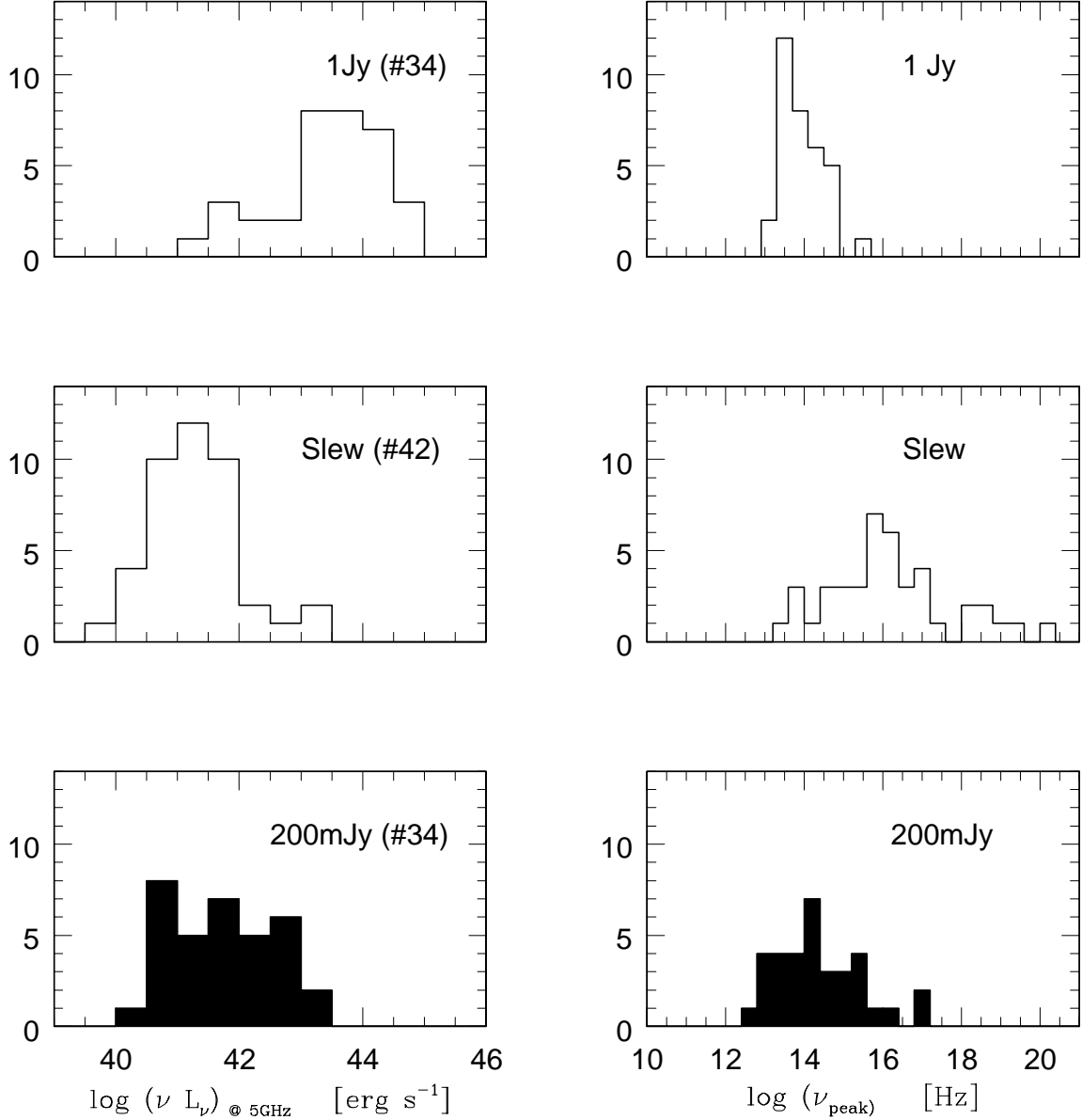}
\caption{For each sample, 1 Jy (top), Slew (middle) and 200 mJy (bottom),
the histograms on the left are the 5 GHz radio luminosity distribution and on 
the right the peak frequency distribution.}
\label{lumpeak}
\end{figure*}

\noindent The 200 mJy being a radio-selected sample, but of relatively
low average radio-luminosity, occupies a critical range of parameter
space and is thus well suited to enable us to check the reality of the
radio-luminosity/spectral sequence: if they follow the expected
luminosity trend, they should have high peak frequencies.  To ensure
that we are comparing like with like, the SEDs of the Slew and 1 Jy
objects listed in Fossati et al (1998) were re-fitted in an
identical manner to the 200 mJy objects
(Section~\ref{sec-sedfitting}). Figure~\ref{lum5nupeak} presents the
peak frequencies against radio luminosities at 5 GHz for the 1 Jy,
Slew and 200 mJy objects (the latter plotted in different symbols
depending on the optical classification). The 200 mJy objects fill a
region of intermediate radio luminosity relative to the 1 Jy and
Slew objects, showing objects with low radio luminosities and low peak
frequencies. This is best shown in Figure~\ref{lumpeak} that presents
the peak frequency distributions of the objects and their respective
radio luminosities at 5 GHz. It is clear that the distributions of
{\it radio luminosities} in the Slew survey and the 200~mJy samples
are very similar. Thus, on the basis of the proposed spectral
sequence, we would expect the 200 mJy objects and Slew objects to have
the same distribution of {\it peak frequencies}. It is, however,
evident from Figure~\ref{lumpeak} that they do not (K-S test rejects
this hypothesis at a significance level of $2.4\times10^{-5}$). The 200~mJy 
peak
frequency distribution is more similar to that of the 1~Jy sample,
suggesting that the method of selection (i.e. radio or X-ray) has more
influence on the peak frequency distribution than does the intrinsic
luminosity.\\

\noindent  We suggest that when
there is a population with a broad distribution of spectral shapes, a
low frequency band of selection biases the samples to objects that are
stronger at low frequencies and a high frequency band of selection
biases the samples to objects that are strong at high
frequencies. Sources strong at low frequencies tend to have low peak
frequencies and visa versa.  In radio astronomical parlance, selecting
at high frequencies is the way to select flat spectrum sources and
selecting at low frequencies produces lots of steep spectrum sources
(Kellermann et al., 1968). Also, as argued in Section~\ref{sec-peak},
there is a bias against recognising as traditional BL Lacs, low
luminosity objects with synchrotron peak frequencies in the infrared
and lower frequencies, which also acts to reinforce the Fossati et
al. correlation.\\

\begin{figure*}
\includegraphics[width=18cm]{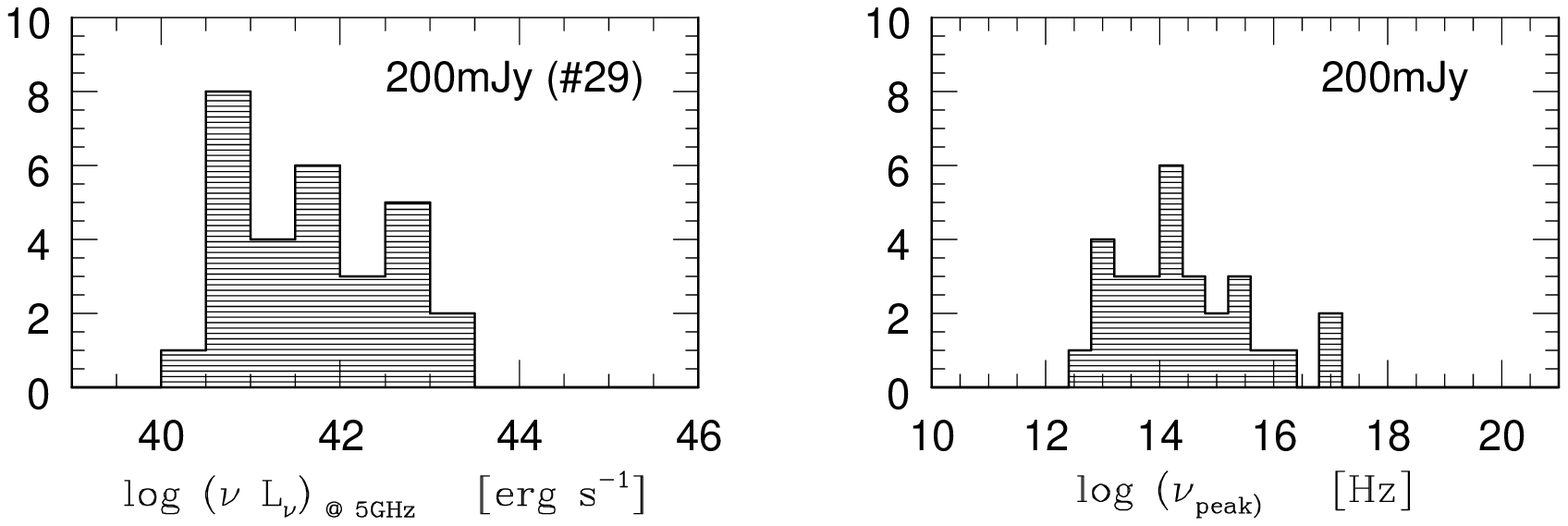}
\caption{The histogram on the left is the 5 GHz radio luminosity distribution 
and on the right the peak frequency distribution of the objects listed in 
Table~\ref{tabela} except the following: 0251+393,
0309+411, 0321+340, 1421+511, 1744+260}
\label{lumpeaknoSy}
\end{figure*}

\section{Discussion and Conclusions}
\label{sec-conclusion}

How valid is our approach to defining blazar samples? The advantage
of ignoring the optical emission is that it is much less subject than
other methods to luminosity-dependent selection effects, something
that is quite important when comparing the statistical properties of
samples with a range of luminosities.\\
\noindent  Perhaps the most controversial
thing we have done is to include objects with Seyfert-like spectra in
our 200 mJy sub-sample. We point out, however, that three of the eight
Seyfert-like objects, namely 0125+487, 1646+466 and 2116+81, are 
well-established blazars (Jackson \& March\~a, 1999; Perlman et al., 1998). 
If we exclude the
remaining five objects (see Figure~\ref{lumpeaknoSy})
it is still clear that the peak
frequency distribution for the 200~mJy sample objects is still much more
like that of the 1Jy sample of high luminosity objects than that of the lower luminosity 
Slew survey objects (K-S test rejects the hypothesis that 200 mJy and Slew
samples have similar peak frequency distributions at a significance level 
of $1.3\times10^{-4}$).\\
\noindent We thus believe that we have shown that amongst low-luminosity radio
sources there are nuclear synchrotron emitters with peaks in their
SEDs falling at lower frequencies than have been found before. These
objects may not meet the criteria usually required for classification
as a blazar but we see no good physical reason why they should be
denied that status and argue that our ``non-optical'' method of sample
selection has significant advantages over the tradition methods. Based
on the assumption that these are really blazars, we have looked at
where these objects fit on the sequence of blazar SEDs with luminosity
investigated by Fossati et al. (1998) and Ghisellini et al. (1998). We
find that they occupy part of the luminosity/peak frequency plane
hitherto almost empty and thus much weaken the significance of the
correlation found by Fossati et al. (1998) and Ghisellini et
al. (1998).  Our results are consistent with the results from Padovani
et al. (2003). We compare peak frequency distributions of samples
selected in diverse ways and argue that the distribution of peak
frequencies for each sample has much more to do with the initial
sample selection frequency than it has to do with intrinsic AGN
luminosity.

\section{Acknowledgments}
 We thank Alessandro Caccianiga and Maria March\~a for useful discussions, 
and Eric Perlman for helpful comments.
S\'onia Ant\'on acknowledges the financial support from the
Portuguese Funda\c c\~ao para a Ci\^encia e Tecnologia through the grant SFRH/BPD/5692/2001,
the European Commission, TMR Programme, Research Network Contract ERBFMRXCT96-0034 ``CERES" and 
the financial support from Jodrell Bank Observatory visitor grant. Ian Browne 
acknowledges the financial support from the
Portuguese Funda\c c\~ao para a Ci\^encia e Tecnologia through the project 
ESO/FNU/43803/2001.

\end{document}